\documentclass[prl,reprint,twocolumn,showpacs,preprintnumbers,amsmath,amssymb,nofootinbib]{revtex4-1}
\usepackage{graphicx}

\usepackage{epsfig}
\usepackage{color}
\usepackage{amssymb}
\usepackage{amsmath}
\usepackage{mathrsfs}
\usepackage{hyperref}
\usepackage{multirow}

\begin{document}

\title{Analytic Two-Loop Higgs Amplitudes in Effective Field Theory \\ and the Maximal Transcendentality Principle}
\author{Qingjun Jin}
\email{qjin@itp.ac.cn}
\author{Gang Yang}
\email{yangg@itp.ac.cn}
\affiliation{CAS Key Laboratory of Theoretical Physics, Institute of Theoretical Physics, \\ Chinese Academy of Sciences, Beijing 100190, China}

\begin{abstract}

We obtain for the first time the two-loop amplitudes for Higgs plus three gluons in Higgs effective field theory including dimension-seven operators. 
This provides the $S$-matrix elements for the top mass corrections for Higgs plus a jet production at LHC. 
The computation is based on the on-shell unitarity method combined with integration by parts reduction. We work in conventional dimensional regularization and obtain analytic expressions renormalized in the modified minimal subtraction scheme. The two-loop anomalous dimensions present operator mixing behavior. The infrared divergences agree with that predicted by Catani and the finite remainders take remarkably simple forms, where the maximally transcendental parts are identical to the corresponding results in ${\cal N}=4$ super-Yang-Mills theory. The parts of lower transcendentality turn out to be also largely determined by the ${\cal N}=4$ results.

\end{abstract}

\maketitle

\noindent \textit{Introduction.}---%
The discovery of a standard-model-like Higgs boson at the Large Hadron Collider (LHC) set a milestone in particle physics. A major goal of the present and future collider experiments is to make precise measurements of the Higgs properties, which is crucial to understanding the precise mechanism of electroweak symmetry breaking and to probe potential new physics beyond the standard model. 

The dominant Higgs production mechanism in the LHC is the gluon fusion through a top quark loop \cite{Ellis:1975ap, Georgi:1977gs}. In the approximation that the top mass $m_{\rm t}$ is much larger than Higgs mass $m_{\rm H}$, the computation can be greatly simplified using an effective field theory (EFT) where the top quark is integrated out \cite{Wilczek:1977zn, Shifman:1979eb, Dawson:1990zj, Djouadi:1991tka, Kniehl:1995tn}. The leading term of the effective Lagrangian, corresponding to taking $m_{\rm t}$ to be infinity, is given by a unique dimension-5 operator, $H{\rm Tr}(G_{\mu\nu}G^{\mu\nu})$, where $H$ is the Higgs field and $G_{\mu\nu}$ is the gauge field strength. 
This provides an accurate approximation for the inclusive Higgs production and has been computed to N$^3$LO QCD accuracy \cite{Anastasiou:2016cez}. 
Differential results for Higgs plus a jet production were also computed at N$^2$LO \cite{Boughezal:2013uia,Chen:2014gva,Boughezal:2015aha,Boughezal:2015dra, Harlander:2016hcx, Anastasiou:2016hlm, Chen:2016zka} in the large $m_{\rm t}$ limit. 
However, if the Higgs transverse momentum is comparable to the top mass, the contribution of the higher dimension operators in the EFT can have significant effect. 
For Higgs plus a jet production, to NLO accuracy, progress of including finite top mass effect was made recently in \cite{Lindert:2018iug,Jones:2018hbb,Neumann:2018bsx}. 
The corresponding planar master integrals were also known analytically in \cite{Bonciani:2016qxi}.

In this Letter, we compute the analytic two-loop amplitudes of Higgs plus three gluons with dimension-7 operators in the EFT. This provides the building blocks for the correction of the top mass effect for Higgs plus a jet production at N$^2$LO order.  We obtain also the two-loop anomalous dimensions of the dimension-7 operators.

Analytic results are crucial for uncovering hidden structures of the amplitudes. 
In this respect, the results of this Letter allow us to test the ``maximal transcendentality principle" which conjectures an intriguing correspondence between QCD and the maximally supersymmetric Yang-Mills (${\cal N}=4$ SYM) theory. 
Here, ``transcendentality" refers to transcendentality weight which is a notion to characterize the ``complexity" of numbers or functions; e.g. the Riemann zeta value $\zeta_n$ or polylogrithm function ${\rm Li}_n$ has weight $n$.
It was first observed in \cite{Kotikov:2002ab, Kotikov:2004er} that, for the anomalous dimensions of twist-two operators, the ${\cal N}=4$ SYM results can be obtained from the maximally transcendental part of the QCD results  \cite{Moch:2004pa}. 
A further surprising observation in \cite{Brandhuber:2012vm} is that the two-loop form factor of stress-tensor multiplet in ${\cal N}=4$ coincides with the maximally transcendental part of the QCD Higgs plus three-gluon amplitudes in the heavy top limit \cite{Gehrmann:2011aa}. The same maximally transcendental function was also found in the ${\cal N}=4$ two-loop Konishi form factor \cite{Banerjee:2016kri}.
Further evidence of this correspondence was found for certain Wilson lines \cite{Li:2014afw, Li:2016ctv}. 
This principle has recently been used to extract from known QCD data the analytic planar four-loop collinear anomalous dimension in ${\cal N}=4$ SYM \cite{Dixon:2017nat}.
While counterexamples of this principle are known such as one-loop amplitudes (see also the study of high energy limit of amplitudes \cite{DelDuca:2017peo}), it is interesting to explore to what extent this principle is valid. 

Our results provide new examples to test the maximal transcendentality principle. 
The two-loop three-gluon form factor of the ${\rm Tr}(G^3)$ operator in ${\cal N}=4$ SYM was obtained in \cite{Brandhuber:2017bkg}, in which it was also argued that the maximally transcendental part should be equal between the ${\cal N}=4$ and QCD counterpart. Our results confirm this argument. More intriguingly, the subleading transcendental parts in QCD turn out to be also closely related to the ${\cal N}=4$ result. In particular, except the transcendental weight zero part, all terms having rational kinematics coefficients are identical between the two theories.

The computation of two-loop amplitudes in QCD, as is well known, is a challenging problem. 
While the analytical two-loop four-gluon amplitudes have been known for a long time \cite{Bern:2000dn, Glover:2001af, Bern:2002tk}, the computation of planar two-loop five-gluon amplitudes is still in progress \cite{Badger:2013gxa, Badger:2015lda, Gehrmann:2015bfy, Dunbar:2016aux, Dunbar:2016cxp, Dunbar:2016gjb, Dunbar:2017nfy,Badger:2017jhb, Abreu:2017hqn,Boels:2018nrr}.
The computation of Higgs amplitudes has extra complications. The inclusion of higher dimension operators introduces new complex interaction vertices and also increases the powers of loop momenta in the integral numerators. 
Furthermore, since the Higgs boson is a color singlet, one encounters nonplanar integrals even for planar Higgs amplitudes, which makes the reconstruction of full integrand via on-shell unitarity method \cite{Bern:1994zx, Bern:1994cg, Britto:2004nc}  highly nontrivial. 

In this Letter, we develop an efficient approach to compute Higgs amplitudes by combining the unitarity method and the integration by parts (IBP) reduction \cite{Chetyrkin:1981qh, Tkachov:1981wb} in an ``unconventional" way.
In particular, we apply the IBP reduction directly for the cut integrands. This avoids reconstructing the full integrand and computes the final coefficients of master integrals. 
Besides, the IBP reduction -- often the most time consuming part of the calculation -- is greatly simplified using the on-shell condition.
A similar idea of combining unitarity cut and IBP reduction has also been used in \cite{Boels:2018nrr}, 
see also \cite{Kosower:2011ty, Larsen:2015ped, Ita:2015tya, Georgoudis:2016wff, Abreu:2017xsl, Abreu:2017hqn}.

\vskip 0.2cm
\noindent \textit{Setup.}---%
Higgs production from gluon fusion can be computed using an effective Lagrangian
\begin{equation}
\label{eq:EFT}
{\cal L}_{\rm eff} = \hat{C}_0 O_0 + {1\over m_{\rm t}^2} \sum_{i=1}^4 \hat{C}_i O_i + {\cal O}\left( {1\over m_{\rm t}^4} \right) \,,
\end{equation}
where $\hat{C}_i$ are Wilson coefficients,  $O_0 = H{\rm Tr}(G^2)$. (Note that $O_0$ also contributes to the ${\cal O}({1\over m_{\rm t}^2})$ order of full physical result  through ${\cal O}({1\over m_{\rm t}^2})$ corrections to the Wilson coefficient $\hat{C}_0$.)
and the subleading terms contain dimension-7 operators \cite{Buchmuller:1985jz, Gracey:2002he, Neill:2009tn, Harlander:2013oja, Dawson:2014ora}
\begin{align}
O_{1} & = H{\rm Tr}(G_\mu^{~\nu} G_\nu^{~\rho} G_\rho^{~\mu}) \,,\\
O_{2} & = H{\rm Tr}(D_\rho G_{\mu\nu} D^\rho G^{\mu\nu} ) \,, \\ 
O_{3} & = H{\rm Tr}(D^\rho G_{\rho\mu} D_\sigma G^{\sigma\mu}) \,,\\
O_{4} & = H{\rm Tr}(G_{\mu\rho} D^\rho D_\sigma G^{\sigma\mu}) \,.
\end{align}
In this paper, we focus on the pure gluon sector. The last two operators have zero contribution in the sector and can contribute when there are internal quark lines, see e.g. \cite{Dawson:2014ora}.

An amplitude with a Higgs boson and $n$ gluons is equivalent to a form factor with the operator ${\cal O}_i$
\begin{equation}
{\cal F}_{{\cal O}_i,n} = \int d^4 x \, e^{-i q\cdot x} \langle p_1, \ldots, p_n | {\cal O}_i(x) |0 \rangle \,,
\end{equation}
where the operator ${\cal O}_i$ is related to a Higgs-gluon interaction term $O_i$ in \eqref{eq:EFT} by $O_i = H {\cal O}_i$ and  $q^2 = m_H^2$.
In the following, we also refer to Higgs amplitudes as form factors.

Using Bianchi identity one has (see e.g. \cite{Gracey:2002he})
\begin{align}
{\cal O}_2 = {1\over2}\, \partial^2{\cal O}_{0} -4\, g_{\rm YM} \, {\cal O}_1 +2\, {\cal O}_{4} \,.
\end{align}
In the pure gluon sector the form factor of ${\cal O}_{4}$ is zero, and we have the relation
\begin{align}
\label{eq:O1-linear-relation}
F_{{\cal O}_2} = {1\over2}\, q^2 \,F_{{\cal O}_{0}} -4\, g_{\rm YM} \, F_{{\cal O}_1} \,.
\end{align}
This will serve as a self-consistency check for the results. 

A simplification of the computation is that for the form factors with three gluons, the color factors factorize out as
\begin{align}
{\cal F}^{(l)}(1_{a_1},2_{a_2},3_{a_3}) = f^{a_1 a_2 a_3} N_c^l F^{(l)}(1,2,3) 
\label{eq:FF_colorstrip}
\end{align}
for $l\leq2$, where $f^{a_1 a_2 a_3}$ is the structure constant of the gauge group. This can be seen by examining the color factors of various two-loop topologies. This implies that the form factor has only planar contribution. Below we consider only the color stripped form factor $F^{(l)}(1,2,3)$.

\vskip 0.2cm
\noindent \textit{Computation.}---%
Unitarity method is a powerful tool to construct loop amplitudes or form factors from their discontinuities, i.e. by applying cuts. On the cut, the loop integrand factorizes into a product of tree-level or lower-loop results. 
While unitarity method is commonly used to reconstruct the full integrand \cite{Bern:1994zx, Bern:1994cg, Britto:2004nc},
we use a novel strategy where the IBP reduction is applied directly to the cut integrand given by the tree products:
\begin{equation}
F^{(l)} |_{\rm cut} = \sum_{\rm helicities} F^{\rm tree} \prod_j A_j^{\rm tree} \stackrel{\textrm{IBP}}{=} \sum_i c_{i} \, M_i |_{\rm cut}  \,.
\end{equation}
In particular, this avoids constructing the full integrand -- which is particularly nontrivial for form factors involving nonplanar diagrams, and one obtains directly the final coefficients $c_i$ of IBP master integrals $M_i$.
Note that a coefficient $c_i$ computed in a single cut channel must be the final answer. 
We would like to stress that our strategy also greatly simplifies the IBP reduction. 
First, the cut integrand is much simpler than the full integrand. Furthermore, the integrals without the cut propagators are dropped off during the reduction.
Below we describe our strategy in more detail.

We apply the $D$-dimensional unitarity method. Tree amplitudes and form factors valid in $D$ dimensions can be computed using color-stripped Feynman diagrams, or recursive techniques such as the Berends-Giele method \cite{Berends:1987me}.
To sum over all helicity states for the cut legs, we contract the internal gluon polarization vectors using:
\begin{equation}
\sum_{\rm helicities}\varepsilon_i^{\mu}\varepsilon_i^{\nu}=\eta^{\mu\nu}
-\frac{q^{\mu}p_i^{\nu}+q^{\nu}p_i^{\mu}}{q\cdot p_i} \,,
\label{eq:helicity-contraction-rule}
\end{equation}
where $q^\mu$ is an arbitrary null momentum. 

Since the cut integrand is gauge invariant, we can further expand it using a set of gauge invariant bases $B_{\alpha}$ (see e.g. \cite{Gehrmann:2011aa, Boels:2017gyc, Boels:2018nrr})
\begin{equation}
{F}_{n}(\varepsilon_i,p_i,l_a)|_{\rm cut}=\sum_{\alpha} {f}^{\alpha}_{n}(p_i, l_a)B_{\alpha}  \,.
\end{equation}
After expansion, all polarization vectors are contained in the basis $B_\alpha$, and $f^{\alpha}_n$ contain only scalar product of loop and external momenta, which can be reduced directly by IBP, using, e.g., public codes \cite{Smirnov:2014hma, Lee:2013mka, vonManteuffel:2012np, Maierhoefer:2017hyi}.  More details are given in the Supplemental Material \cite{supplemental}.

\begin{figure}[t]
\centering
\includegraphics[clip,scale=0.28]{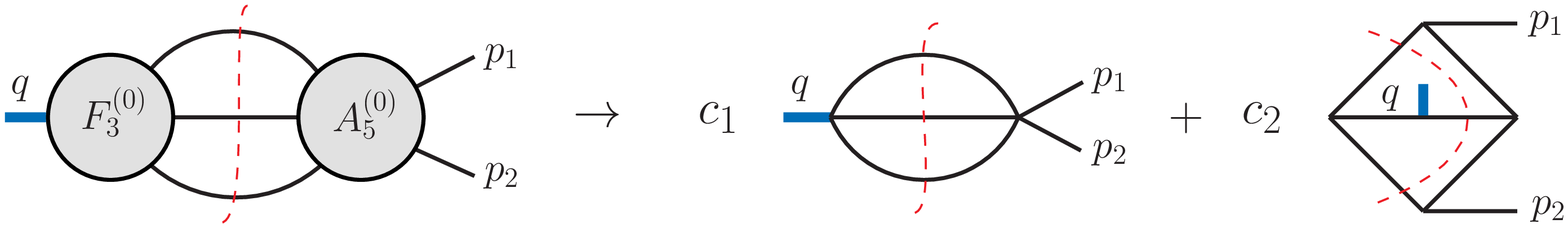}
\caption[a]{Triple cut for two-loop form factor.}
\label{fig:triplecutExample}
\end{figure}

\begin{figure}[t]
\centering
\includegraphics[clip,scale=0.23]{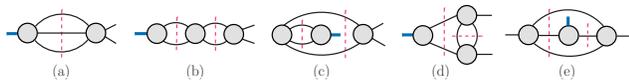}
\caption[a]{Cuts needed for the 2-point form factors.}
\label{fig:FF_2g_2loop_Allcuts}
\end{figure}

As an example, consider the triple cut of $F_{{\cal O}_2}^{(2)}(p_1,p_2)$ in Fig \ref{fig:triplecutExample}. 
Starting from the tree products $F_3^{(0)}  \cdot A_5^{(0)}$ and using the procedure described above, this cut allows us to fix the coefficients of the sunrise and the cross-ladder integrals. 
To determine the coefficients of all master integrals, there are four other cuts to consider, as shown in Fig \ref{fig:FF_2g_2loop_Allcuts}.

\begin{figure}[t]
\centering
\includegraphics[clip,scale=0.245]{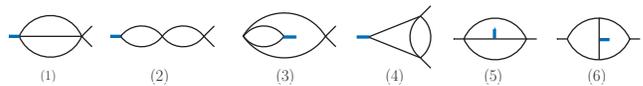}
\caption[a]{Master integrals of the 2-loop 2-point form factor.}
\label{fig:FF_2g_2loop_MIs}
\end{figure}

Let us explain an important new feature of form factors compared to amplitudes computation. Since the operator (i.e., Higgs particle) is a color singlet, the Higgs leg can appear in the ``internal" part of the graph, even for the color-planar contribution. This explains the appearance of the nonplanar cuts (c) and (e) in Fig. \ref{fig:FF_2g_2loop_Allcuts}, which determine the coefficients of master integrals (3) and (5), respectively, in Fig. \ref{fig:FF_2g_2loop_MIs}. Although integrals (3) [and (5)] are mathematically equivalent to integrals (2) [and (4)], they have different physical origin and should be considered separately.
The full form factor $F_{{\cal O}_2}^{(2)}$ is given as
\begin{equation}
F_{{\cal O}_2}^{(2)}(p_1,p_2) = \bigg( \sum_{i=1}^4 c_i M_i + \sum_{i=5,6} {c_i\over2} M_i \bigg) + \textrm{perms}(p_1, p_2) ,
\label{eq:FF_O1_2pt}
\end{equation}
where $M_i$ correspond to the integrals with label $(i)$, $i=1,\ldots,6$, in Fig \ref{fig:FF_2g_2loop_MIs}. Note the factor $\frac{1}{2}$ is necessary for integrals (5) and (6), since the permutation does not alter the diagram.

\begin{figure}[t]
\centering
\includegraphics[clip,scale=0.26]{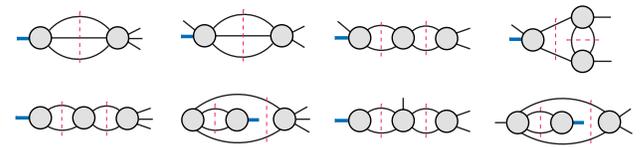}
\caption[a]{Cuts needed for the three-point form factors.}
\label{fig:FF_3g_2loop_Allcuts}
\end{figure}

\begin{figure}[t]
\centering
\includegraphics[clip,scale=0.23]{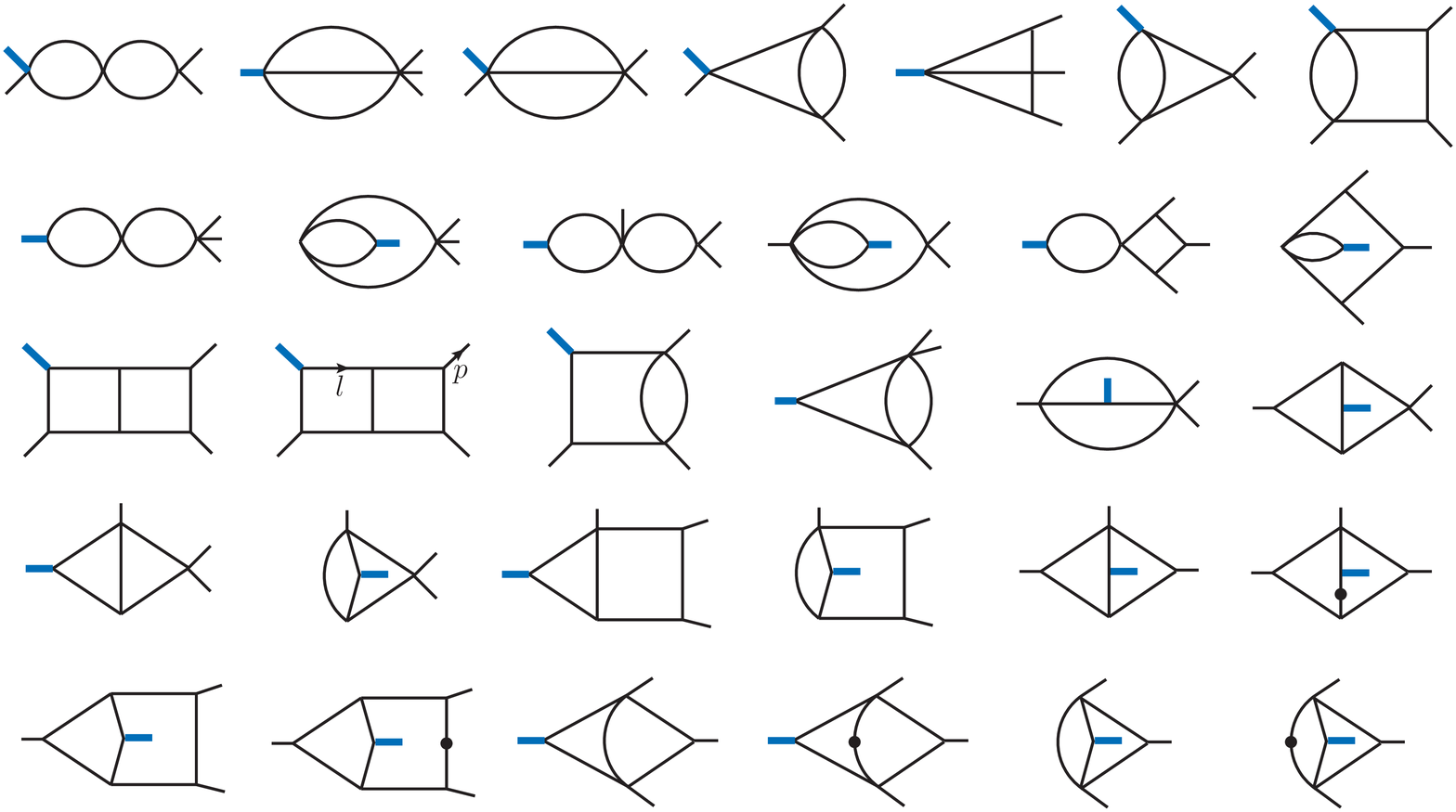}
\caption[a]{Master integrals of the two-loop three-point form factors.}
\label{fig:FF_3g_2loop_MIs}
\end{figure}

For the three-point two-loop form factors, all the cuts needed are given in Fig \ref{fig:FF_3g_2loop_Allcuts}. 
The master integrals are shown in Fig \ref{fig:FF_3g_2loop_MIs}.
While all cuts are needed for the form factor of length-2 operator ${\cal O}_0$ and ${\cal O}_2$, only the first four cuts contribute to ${\cal O}_1$, since the tree form factors of ${\cal O}_1$ contain at least three gluons. Accordingly, only the first seven master integrals in Fig \ref{fig:FF_3g_2loop_MIs} appear in the form factor of ${\cal O}_1$. 
The full form factor is obtained by adding all the master integrals and taking into account the symmetry factors properly, similar to the two-point case in \eqref{eq:FF_O1_2pt}.

We compute all two-loop form factors of ${\cal O}_i, i=0,1,2$ with two and three external gluons. 
We would like to emphasize that the computation of the form factor for ${\cal O}_2$ is more involved than the known result of ${\cal O}_{0}$ due to extra derivatives in the operator. 

The above strategy can be also applied to ${\cal N}=4$ SYM. One can use four-dimensional helicity tree amplitudes and form factors in the cuts, which corresponds to the use of the four-dimensional helicity (FDH) scheme. 
With this strategy we obtain the ${\cal N}=4$ form factors of the super extension of ${\cal O}_0$ and ${\cal O}_1$ previously computed in \cite{Brandhuber:2012vm} and \cite{Brandhuber:2017bkg}.

We provide explicit IBP coefficients of the form factor of ${\cal O}_1$ and its ${\cal N}=4$ counterpart in the Supplemental Material \cite{supplemental}.

\vskip 0.2cm
\noindent \textit{Divergence subtraction and checks.}---%
The bare form factors contain both ultraviolet (UV) and infrared (IR) divergences. 
The $\epsilon$ expansion of the bare form factors can be obtained using  \cite{Gehrmann:2000zt,Gehrmann:2001ck}, where all master integrals were computed.
Our QCD results are regularized in the CDR scheme, and we use the modified minimal subtraction renormalization scheme \cite{Bardeen:1978yd}. To remove the UV divergences, both the gauge coupling and the operator require renormalization. For the IR divergences, we apply the subtraction formula by Catani  \cite{Catani:1998bh}. 

At two loops, all poles in $1/\epsilon^m, m=4,3,2$ are totally fixed by the universal IR structure and the one-loop data. 
The $1/\epsilon$ UV poles contain the information of two-loop anomalous dimensions, which are related to the renormalization constants of the operators as 
\begin{align}
\label{eq:AD_from_RC}
\gamma = \mu \frac{\partial}{\partial \mu} \log Z \,. 
\end{align}

Our computations reproduce all known results, including the nontrivial two-loop QCD amplitudes of Higgs plus three gluons with the operator ${\cal O}_{0}$ \cite{Gehrmann:2011aa} (see also \cite{Duhr:2012fh}). We match not only the divergences but also the finite remainders exactly. Our ${\cal N}=4$ computations also reproduce the results in \cite{Brandhuber:2012vm} and \cite{Brandhuber:2017bkg}.

As a further consistency check of the new results of dimension-7 operators, 
we find they satisfy exactly the linear relation \eqref{eq:O1-linear-relation}. This is true already for the expressions in terms of IBP master integrals.

\vskip 0.2cm
\noindent \textit{Operator mixing at two loops.}---%
At two loops the operator mixing appears.
Let us first consider ${\cal O}_2$. Based on \eqref{eq:O1-linear-relation}, we can define a new operator
\begin{equation}
\tilde{\cal O}_2 = -\frac{3}{2}( {\cal O}_2 + 8 g_{\rm YM} \, {\cal O}_1 ) = -\frac{3}{4}\, \partial^2{\cal O}_{0} \,,
\label{eq:def_tildeO1}
\end{equation}
which has no mixing with others. 
The form factor of $\tilde{\cal O}_2$ is proportional to that of ${\cal O}_0$ as
$F_{\tilde{\cal O}_2} = -\frac{3}{4}\, q^2 \,F_{{\cal O}_0}$.
The normalization constant $-\frac{3}{4}$ is introduced such that 
${F_{\tilde{\cal O}_2}^{(0)}(1^-,2^-,3^-) / F_{{\cal O}_1}^{(0)}(1^-,2^-,3^-)} = {1/ (u v w)}$,
where
\begin{align}
u = {s_{12} \over q^2 } \,, \quad v = {s_{23} \over q^2} \,, \quad w = {s_{13} \over q^2} \,, \qquad q^2 = s_{123} \,.
\end{align}

To study the operator mixing effect for ${\cal O}_1$, we first consider the form factor with two external gluons  $F_{{\cal O}_1}^{(l)}(1^-,2^-)$. The tree and one-loop results are zero, while at two loops we obtain
\begin{equation}
F_{{\cal O}_1}^{(2)}(1,2) = F_{\tilde{\cal O}_2}^{(0)}(1,2) \left( -{1\over \epsilon} + 2 \log s_{12} - \frac{487}{72} \right) + {\cal O}(\epsilon) \,.
\label{eq:FO2-2g-Z2}
\end{equation}
This is completely an operator mixing effect between ${\cal O}_1$ and $\tilde{\cal O}_2$. 
For the three-point case $F_{{\cal O}_1}^{(2)}(1^-,2^-,3^-)$, the part related to renormalization constant $Z^{(2)}$ is given as
\begin{align}
\label{eq:FO2-3gluon-Z2}
& F_{{\cal O}_1}^{(2)}(1^-,2^-,3^-) \big|_{Z^{(2)}\textrm{-part}} \\ 
& = F_{{\cal O}_1}^{(0)}(1^-,2^-,3^-) \left( -{19\over 24 \epsilon^2} + {25\over 12 \epsilon}  - {1 \over u v w} {1\over\epsilon} \right) \,. \nonumber
\end{align}
The term ${1 \over u v w}$ is precisely due to the operator mixing,  and its divergence is consistent with \eqref{eq:FO2-2g-Z2}.

We can define a new operator to avoid the operator mixing as
\begin{align}
{\tilde{\cal O}}_1 &= {\cal O}_1 + {1\over \epsilon}{1\over g_{\rm YM}} \left({\alpha_s \over 4\pi}\right)^2 \tilde{\cal O}_2 \,,
\end{align}
and from \eqref{eq:FO2-3gluon-Z2} we have 
\begin{align}
Z_{{\tilde{\cal O}}_1}^{(2)} = -{19\over 24 \epsilon^2} + {25\over 12 \epsilon}  \,, \qquad 
\gamma_{\tilde{\cal O}_1}^{(2)} = {25\over3}  \,,
\label{eq:ZO2-2loop}
\end{align}
where $\gamma_{\tilde{\cal O}_1}^{(2)}$ is computed using \eqref{eq:AD_from_RC}.

\vskip 0.2cm
\noindent \textit{Two-loop finite remainder.}---%
After renormalization and subtracting the IR divergences, the two-loop finite remainder of $F_{{\rm R},{\cal O}_1}^{(2)}(1^-,2^-,3^-)$ is given in terms of two-dimensional harmonic polylogarithms \cite{Gehrmann:2000zt, Gehrmann:2001jv}, which can be simplified using the symbology technique \cite{Goncharov:2010jf}. The final expression takes a remarkable simple form. It consists of functions of transcendentality weight ranging from 4 to 0 and can be decomposed as:
\begin{align}
F_{{\rm R},{\cal O}_1}^{(2),{\rm fin}} = F_{{\cal O}_1}^{(0)} \sum_{i=0}^4 \Omega^{(2)}_{{\cal O}_1; i} ,
\end{align}
where $\Omega^{(2)}_{{\cal O}_1;i}$ has uniform transcendentality weight $i$.
To properly compare with the ${\cal N}=4$ form factor, we compute the latter in the Catani subtraction scheme \cite{Catani:1998bh}, denoted by $\Omega^{(2), {\cal N}=4}_{{\cal O}_1;i}$. This is different from the result in \cite{Brandhuber:2017bkg} based on the BDS subtraction scheme \cite{Bern:2005iz}.  

Below we give the explicit QCD results according the transcendentality weight and comment on their relation to the corresponding ${\cal N}=4$ counterparts.
As we will see, not only the maximally transcendental parts are identical, the lower transcendental parts are also closely related to each other.

Weight 4: The maximally transcendental part is given by:
\begin{align}
\Omega^{(2)}_{{\cal O}_1;4} = & -{3\over2} {\rm Li}_4(u) + {3\over4} {\rm Li}_4\left(-{u v \over w} \right) - {3\over2} \log(w)  {\rm Li}_3 \left(-{u\over v} \right)\nonumber\\
& + {\log^2(u) \over 32} \left[ \log^2(u) + \log^2(v) \right. \nonumber\\
& \left.+ \log^2(w) - 4\log(v)\log(w) \right] \nonumber\\
& + {\zeta_2 \over 8} \left[ 5\log^2(u) - 2 \log(v)\log(w) \right]- {1\over4} \zeta_4 \nonumber\\
&- {1\over2} \zeta_3 \log(-q^2) + \textrm{perms}(u,v,w) \,.
\end{align}
We find a precise match between QCD and ${\cal N}=4$ results:
$\Omega^{(2)}_{{\cal O}_1;4} = \Omega^{(2), {\cal N}=4}_{{\cal O}_1;4}$,
which confirms the argument made in \cite{Brandhuber:2017bkg}. Note that the expression is slightly different from the result of \cite{Brandhuber:2017bkg}, which also appears in other form factors in ${\cal N}=4$ SYM \cite{Brandhuber:2014ica, Loebbert:2015ova, Brandhuber:2016fni, Loebbert:2016xkw}; this is purely due to the change of scheme between Catani and BDS subtraction.

Weight 3: The transcendentality-3 part is given by:
\begin{align}
\Omega^{(2)}_{{\cal O}_1;3} = &  \left( 1+ {u\over w} \right) T_3  + {143\over72}\zeta_3 - {11\over24}\zeta_2\log(-u \,q^2) 
\nonumber\\ & + \textrm{perms}(u,v,w) \,,
\end{align}
where 
\begin{align}
T_3 := & \Big[ -{\rm Li}_3 \left(-{u\over w} \right) + \log(u) {\rm Li}_2\left({v \over 1-u} \right) \nonumber\\
& - {1\over2} \log(1-u) \log(u) \log\left({w^2\over 1-u}\right)  + {1\over2} {\rm Li}_3\left(-{uv \over w}\right)  \nonumber\\
& + {1\over2} \log(u)\log(v)\log(w) + {1\over12}\log^3(w) + (u\leftrightarrow v) \Big] \nonumber\\
&+  {\rm Li}_3(1-v) - {\rm Li}_3(u) + {1\over2} \log^2(v) \log\left({1-v\over u}\right) \nonumber\\
&- \zeta_2 \log\left( {u v \over w} \right) \,.
\end{align}
Very interestingly, the corresponding ${\cal N}=4$ SYM result is given by
\begin{align}
\Omega^{(2), {\cal N}=4}_{{\cal O}_1;3} = & \left( 1+ {u\over w} \right) T_3 + \textrm{perms}(u,v,w) \,.
\end{align}
The function $T_3$ also appeared as the building block of the corresponding  ${\cal N}=4$ result \cite{Brandhuber:2017bkg} and in the form factor in the SU(2) sector in ${\cal N}=4$ \cite{Loebbert:2015ova}.

Weight 2: The transcendentality-2 part is given by:
\begin{align}
\Omega^{(2)}_{{\cal O}_1;2} = & \bigg\{ \left( {u^2\over w^2} + {v^2\over w^2} -1 \right) \Big[{\rm Li}_2(1-u) + {1\over2} \log(u) \log(v) \nonumber\\
&   - {1\over2} \zeta_2 \Big] - {55 \over 48}\log^2(u) + {73\over 72}\log(u) \log(v)  + {23\over6} \zeta_2 \nonumber\\
& + \textrm{perms}(u,v,w) \bigg\}  - {19\over36}\log(u v w) \log(-q^2) \nonumber\\
& - {19\over24} \log^2(-q^2)  \,,
\end{align}
where the terms containing the rational  coefficient $u^2/w^2+ v^2/w^2$ is identical to the ${\cal N}=4$ result.

Weight 1 and 0: The weight-1 part is given by:
\begin{align}
\Omega^{(2)}_{{\cal O}_1;1}  = & \left( {119\over18} + {v\over w} + { u^2 \over 2 v w} \right) \log(u) \\
& + \left( {119\over18} - { 1 \over 3 u v w} \right) \log(-q^2)  + \textrm{perms}(u,v,w) \,, \nonumber
\end{align}
where the terms with coefficients that are rational functions of $\{u,v,w\}$ are identical to the  ${\cal N}=4$ result.
Finally, the weight-0 part is given by:
\begin{align}
\Omega^{(2)}_{{\cal O}_1;0} = {487\over 72}{1\over u v w} - {14075\over 216}  \,,
\end{align}
where we note that the coefficient of ${1\over u v w}$ equals the finite rational number in \eqref{eq:FO2-2g-Z2}.

\vskip .2cm
\noindent \textit{Summary and discussion.}---%
We obtain the first analytic two-loop Higgs amplitudes with dimension-7 operators in the Higgs EFT.
These contribute to the top mass corrections for Higgs plus a jet production at N$^2$LO order. 
Our computation is based on an efficient new method of combining the on-shell unitarity method and IBP reduction,
which can be also applied to dimension-9 operators and beyond.
The final results are given in complete analytic form which reveal direct connections between QCD and ${\cal N}=4$ SYM. The maximally transcendental part of the form factor of ${\rm Tr}(G^3)$ turns out to be equivalent to that of ${\cal N}=4$ SYM, as was argued in \cite{Brandhuber:2017bkg, Brandhuber:2018xzk} (see also \cite{Brandhuber:2018kqb} for the study of ${\cal N}<4$ supersymmetric theories). 
More intriguingly, we find the parts of lower transcendentalities are also similar to the ${\cal N}=4$ blocks. In particular, except the transcendental weight zero term, all terms having coefficients of rational functions of $\{u,v,w\}$ are identical between two theories.
Furthermore, because of the linear relation \eqref{eq:O1-linear-relation}, this implies that the maximal transcendentality principle applies also to the form factors of  ${\rm Tr}(DGDG)$ operator and the corresponding supermultiplet in ${\cal N}=4$ SYM. 
The surprising simplicity of the results indicates that there should be alternative path to understand or derive the results in a more direct way, e.g. using bootstrap method (see e.g. \cite{Caron-Huot:2016owq, Dixon:2016nkn, Almelid:2017qju}).

\vskip 0.1cm

It is a pleasure to thank Rutger Boels, Andi Brandhuber, Hui Luo, Jianping Ma, Ming Yu, Huaxing Zhu, and in particular Lance Dixon for discussions. 
This work is supported by the Chinese Academy of Sciences (CAS) Hundred-Talent Program, by the HPC Cluster of ITP-CAS, by the Key Research Program of Frontier Sciences of CAS, and by Project 11747601 supported by National Natural Science Foundation of China.

\bibliographystyle{apsrev4-1}


%

\newpage

\onecolumngrid
\newpage
\appendix

\section*{Supplemental material}

In this supplemental material, we first give some details of the Lorentz invariant basis expansion, then we provide the explicit two-loop form factor of  ${\rm Tr}(G^3)$  in terms of IBP master integrals as well as its comparison with its ${\cal N}=4$ counterpart.

\subsection{Lorentz invariant basis expansion}

Since the cut-integrand is gauge invariant, we can expand the integral using a set of gauge invariant basis $B_{\alpha}$ (see e.g. \cite{Gehrmann:2011aa, Boels:2017gyc, Boels:2018nrr})
\begin{equation}
{F}_{n}(\varepsilon_i,p_i,l_a)|_{\rm cut}=\sum_{\alpha} {f}^{\alpha}_{n}(p_i, l_a)B_{\alpha}  \,.
\label{eq:F-expand}
\end{equation}
For the form factor with three external gluons, the gauge invariant basis has four elements and can be chosen explicitly as  
\begin{equation}
B_1= A_1 C_{23}\,,\quad B_2= A_2 C_{31} \,,\quad B_3= A_3 C_{12} \,,\quad B_4=A_1A_2A_3 \,,
\end{equation}
in which $A_{i}$ and $C_{ij}$ are defined by
\begin{equation}
A_{i} = \frac{\varepsilon_i \cdot p_j }{p_i\cdot p_j} -\frac{\varepsilon_i\cdot p_k}{p_i\cdot p_k} \,, \quad
C_{ij} = \varepsilon_i\cdot \varepsilon_j -\frac{(p_i\cdot \varepsilon_j)(p_j\cdot \varepsilon_i)}{ p_i\cdot p_j} \,,
\end{equation}
where the $\{i,j,k\}$ in $A_{i}$ are cyclic permutations of $\{1,2,3\}$. For form factors with only two external gluons, there is only one gauge basis $B_0=C_{12}$.

To provide more details on the expansion, we first recall the contract rule of  polarization vectors:
\begin{equation}
\varepsilon_i^{\mu} \circ \varepsilon_i^{\nu}\equiv \sum_{\rm helicities}\varepsilon_i^{\mu}\varepsilon_i^{\nu}=\eta^{\mu\nu}
-\frac{q^{\mu}p_i^{\nu}+q^{\nu}p_i^{\mu}}{q\cdot p_i} \,,
\label{eq:helicity-contraction-rule}
\end{equation}
where $q^\mu$ is an arbitrary lightlike reference momenta. 
Then the coefficient ${f}^{\alpha}_{n}(p_i, l_a)$ in \eqref{eq:F-expand} can be computed as
\begin{equation}
{f}^{\alpha}_{n}(p_i, l_a)=B^{\alpha}\circ{F}_{n}(\varepsilon_i,p_i,l_a) \,,
\end{equation}
where $B^{\alpha}$ are the dual basis and defined as
\begin{equation}
B^{\alpha}\circ B_{\beta}=\delta^{\alpha}_{\beta},\ 
B_{\alpha}=G_{\alpha\beta}B^{\beta},\ 
G_{\alpha\beta}=B_{\alpha}\circ B_{\beta}\ .
\end{equation}

After expansion, all the polarization vectors $\varepsilon_i$ are contained in the basis $B_\alpha$, and ${f}^{\alpha}_n$ contain only scalar product of loop and external momenta, which can be reduced directly using IBP (with cut conditions) as
\begin{equation}
{f}^{\alpha}_{n}= \sum_i c_i^\alpha \, M_i |_{\rm cut} \,,
\end{equation}
where $M_i$ are IBP master integrals. The final form factor in a given cut is thus expanded in terms of master integrals as
\begin{equation}
F^{(l)} |_{\rm cut} =  \sum_i c_{i} \, M_i |_{\rm cut} =  \sum_{i,\alpha} (c_i^\alpha B_\alpha) \, M_i |_{\rm cut}  \,.
\end{equation}

\newpage
\subsection{Two-loop form factor of ${\rm Tr}(G^3)$}

\begin{figure}[ht]
\centering
\includegraphics[clip,scale=0.4]{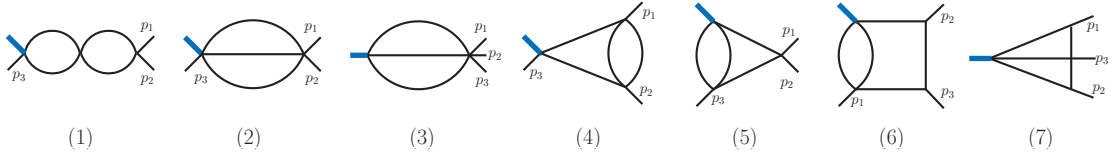}
\caption[a]{The master integrals of the 2-loop 2-point form factor.}
\label{fig:FFF_3g_2loop_MIs}
\end{figure}

The bare two-loop form factor of ${\rm Tr}(G^3)$ form factor can be expressed in terms of master integrals as
\begin{align}
F^{(2)} = c_{1} I_{1} + c_2 I_2 + {1\over3} c_3 I_3 + c_4 I_4 + c_5 I_5 + c_6 I_6 + (c_6 I_6) |_{p_2 \leftrightarrow p_3}  + c_7 I_7 + \textrm{cyclic perms}(p_1,p_2,p_3) \,,
\end{align} 
where the master integrals $I_\alpha$ are shown in Fig.~\ref{fig:FFF_3g_2loop_MIs}, and the subscripts $\alpha$ correspond to the labels of the figures. The expressions of master integrals in terms of harmonic polylogarithms can be found in \cite{Gehrmann:2000zt}. 

Below we give the coefficients of master integrals:
\begin{align}
c_1^{\rm QCD} = & \frac{1}{\epsilon ^2}-\frac{15}{4 (2 \epsilon -3)}+\frac{81}{8 (2 \epsilon -3)^2}+\frac{3}{\epsilon -1}+\frac{1}{2
   (\epsilon -1)^2}+\frac{1}{8} \,, \\
c_2^{\rm QCD} = &
\frac{9}{s_{12} \epsilon ^3}-\frac{139 s_{13} s_{23}+12 s_{12} 
(s_{13}+s_{23})}{6 s_{12} s_{13} s_{23} \epsilon ^2}
+\frac{361 s_{13} s_{23}+54 s_{12} \
(s_{13}+s_{23})}{18 s_{12} s_{13} s_{23} \epsilon }
\nonumber\\ &
-\frac{12}{s_{12} 
(3-2 \epsilon )^2}
-\frac{s_{12}^2+12 s_{13} s_{23}-s_{12} 
(s_{13}+s_{23})}{12 s_{12} s_{13} s_{23} (\epsilon -1
)^2}
\nonumber\\&
+\left(-\frac{37}{6 s_{12}}+\frac{7}{12 s_{13}}+
\frac{7}{12 s_{23}}+\frac{s_{12}-s_{23}}{6 
s_{13}^2}+\frac{s_{12}-s_{13}}{6 s_{23}^2}-\frac{s_{12}}{4 s_{13} 
s_{23}}\right)\frac{1}{\epsilon -1}
\nonumber\\ &
+\left(\frac{134}{9 s_{12}}-\frac{3 (s_{12}+s_{23})}{2 
s_{13}^2}-\frac{3 (s_{12}+s_{13})}{2 s_{23}^2}+\frac{7 s_{12}}{s_{13} 
s_{23}}+\frac{3 s_{13}}{s_{12} s_{23}}+\frac{3 s_{23}}{s_{12} 
s_{13}}\right)\frac{1}{2 \epsilon -3}
\nonumber\\ &
-\left(\frac{7}{2 \
s_{12}}+\frac{s_{12}+s_{23}}{2 s_{13}^2}+\frac{s_{12}+s_{13}}{2 
s_{23}^2}+\frac{s_{12}}{2 s_{13} s_{23}}+\frac{s_{13}}{2 s_{12} 
s_{23}}+\frac{s_{23}}{2 s_{12} s_{13}}+\frac{2 s_{12} s_{13}}{3 
s_{23}^3}+\frac{2 s_{12} s_{23}}{3 s_{13}^3}\right)\frac{1}{2 \epsilon -1} 
\nonumber\\ &
+\left(-\frac{4}{3 
s_{12}}+\frac{1}{s_{13}}+\frac{1}{s_{23}}-\frac{7 s_{123}}{6 
s_{23}^2}-\frac{7 s_{123}}{6 s_{13}^2}-\frac{2 s_{12} s_{13}}{3 
s_{23}^3}-\frac{2 s_{12} s_{23}}{3 s_{13}^3}+\frac{11 s_{12}}{3 
s_{13} s_{23}}+\frac{5 s_{13}}{2 s_{12} s_{23}}+\frac{5 s_{23}}{2 
s_{12} s_{13}}\right)
\nonumber\\ &
+
\left(\frac{1}{s_{13}}+\frac{1}{s_{23}}+\frac{s_{23}}{s_{13}^2}+\frac{
s_{13}}{s_{23}^2}+\frac{s_{12}}{s_{13} s_{23}}+\frac{s_{13}}{s_{12} 
s_{23}}+\frac{s_{23}}{s_{12} s_{13}}+\frac{s_{12} 
s_{23}}{s_{13}^3}+\frac{s_{12} s_{13}}{s_{23}^3}\right) \epsilon  
\,, \\
c_3^{\rm QCD} = & 
\left(\frac{3}{2 s_{12}}+\frac{3}{2 s_{13}}+
\frac{3}{2 s_{23}}+\frac{s_{123}}{2 s_{12}^2}+\frac{s_{123}}{2 
s_{23}^2}+\frac{s_{123}}{2 s_{13}^2}+\frac{3 s_{12}}{2 s_{13} 
s_{23}}+\frac{3 s_{13}}{2 s_{12} s_{23}}+\frac{3 s_{23}}{2 s_{12} 
s_{13}}+\frac{2 s_{12} s_{23}}{3 s_{13}^3}+\frac{2 s_{12} s_{13}}{3 
s_{23}^3}
\right.
\nonumber\\ &
\left.+\frac{2 s_{13} s_{23}}{3 s_{12}^3}\right)\frac{1}{2 \epsilon -1} 
+\left(-\frac{49}{3 s_{12}}-\frac{49}{3 s_{13}}-\frac{49}{3 
s_{23}}+\frac{s_{123}}{s_{13}^2}+\frac{s_{123}}{s_{12}^2}+\frac{s_{
123}}{s_{23}^2}-\frac{11 s_{12}}{s_{13} s_{23}}-\frac{11 
s_{13}}{s_{12} s_{23}}-\frac{11 s_{23}}{s_{12} s_{13}}
\right.
\nonumber\\ &
\left.+\frac{2 s_{12} 
s_{13}}{3 s_{23}^3}
+\frac{2 s_{12} s_{23}}{3 s_{13}^3}+\frac{2 s_{13} 
s_{23}}{3 s_{12}^3}\right)
+\frac{2 s_{12}^2+2 s_{23}^2+2 
s_{13}^2+s_{123}^2}{12 s_{12} s_{13} s_{23} (\epsilon -1
)^2}+\frac{10 s_{12}^2+10 s_{13}^2+10 s_{23}^2-s_{123}^2}{12 s_{12} 
s_{13} s_{23} (\epsilon -1)}
\nonumber\\ &
+\left(-\frac{57}{2 s_{12}}-\frac{57}{2 s_{13}}-\frac{57}{2 
s_{23}}+\frac{3 s_{123}}{2 s_{12}^2}+\frac{3 s_{123}}{2 
s_{13}^2}+\frac{3 s_{123}}{2 s_{23}^2}-\frac{21 s_{12}}{s_{13} 
s_{23}}-\frac{21 s_{13}}{s_{12} s_{23}}-\frac{21 s_{23}}{s_{12} 
s_{13}}\right)\frac{1}{2 \epsilon -3}
\nonumber\\ &
- 
\left(\frac{5}{s_{12}}+\frac{5}{s_{13}}+\frac{5}{s_{23}}+\frac{3 
s_{12}}{s_{13} s_{23}}+\frac{3 s_{13}}{s_{12} s_{23}}+\frac{3 
s_{23}}{s_{12} s_{13}}+\frac{s_{12} s_{13}}{s_{23}^3}+\frac{s_{12} 
s_{23}}{s_{13}^3}+\frac{s_{13} s_{23}}{s_{12}^3}\right)\epsilon 
\,, \\
c_4^{\rm QCD} = & \frac{3}{2 \epsilon ^2}-\frac{5}{3 \epsilon }-\frac{5}{\epsilon 
-1}+\frac{559}{30 (2 \epsilon -3)}-\frac{28}{15 (3 \epsilon 
-2)}-\frac{1}{(\epsilon -1)^2}+\frac{1}{6} +\frac{\epsilon }{2}\,, %
\end{align}
\begin{align}
c_5^{\rm QCD} = & \frac{\left(s_{13}^2-4 s_{13} s_{23}+s_{23}^2\right) 
(s_{13}+s_{23})^2}{6 s_{13}^2 s_{23}^2 (\epsilon -1)}-\frac{4 \left(5 
s_{13}^2+7 s_{13} s_{23}+5 s_{23}^2\right) (s_{13}+s_{23})^2}{45 
s_{13}^2 s_{23}^2 (3 \epsilon -2)} \nonumber\\ &
-\frac{\left(s_{13}^2-7 s_{13} 
s_{23}+s_{23}^2\right) (s_{13}+s_{23})^2}{18 s_{13}^2 
s_{23}^2}+\frac{(s_{13}+s_{23})^2}{s_{13} s_{23} \epsilon }+\frac{18 
(s_{13}+s_{23})^2}{5 s_{13} s_{23} (2 \epsilon 
-3)}-\frac{(s_{13}+s_{23})^2}{6 s_{13} s_{23} (\epsilon -1)^2} \,,
\\
c_6^{\rm QCD} = & \frac{s_{23}}{\epsilon } +\frac{s_{23} 
\left(-762 s_{12}^2+832 s_{12} s_{13}+35 s_{13}^2\right)}{756 
s_{12}^2 (3 \epsilon -1)}
+ \frac{s_{23} \left(-6 s_{12}^2-4 s_{12} s_{13}+s_{13}^2\right)}{12 
s_{12}^2 (\epsilon -1)}
\nonumber\\ &
+\frac{s_{23} (6 s_{12}-s_{13}) (2 
s_{12}+s_{13})}{54 s_{12}^2}+\frac{18 s_{23} (4 s_{12}+3 s_{13})}{35 
s_{12} (2 \epsilon -3)}
+\frac{4 s_{23} \left(3 s_{12}^2-14 s_{12} 
s_{13}-10 s_{13}^2\right)}{135 s_{12}^2 (3 \epsilon -2)}
 \,, \\
c_7^{\rm QCD} = & 
\frac{3 s_{12}^4+6 s_{12}^3 (s_{13}+s_{23})+3 s_{12}^2 
\left(s_{13}^2+s_{23}^2\right)+3 s_{12} s_{13} s_{23} 
(s_{13}+s_{23})+4 s_{13}^2 s_{23}^2}{6 s_{12}^3 (2 \epsilon 
-1)}
\nonumber\\ &
-\frac{27 \left(2 s_{12}^2 (s_{13}+s_{23})-2 s_{12} \left(s_{13}^2-3 \
s_{13} s_{23}+s_{23}^2\right)+s_{13} s_{23} (s_{13}+s_{23})\right)}{70 s_{12}^2 (2 \epsilon -3)}
\nonumber\\ &
+\frac{6 s_{12}^2 \
(s_{13}+s_{23})+9 s_{12} s_{13} s_{23}-s_{13} s_{23} 
(s_{13}+s_{23})}{12 s_{12}^2 (\epsilon -1)}
\nonumber\\ &
-\frac{4 \left(8 s_{12}^3 (s_{13}+s_{23})+s_{12}^2 \left(2 
s_{13}^2-11 s_{13} s_{23}+2 s_{23}^2\right)+9 s_{12} s_{13} s_{23} 
(s_{13}+s_{23})+30 s_{13}^2 s_{23}^2\right)}{135 s_{12}^3 (3 \epsilon 
-2)}
\nonumber\\ &
-\frac{-252 s_{12}^4+766 s_{12}^3 (s_{13}+s_{23})+s_{12}^2 
\left(256 s_{13}^2+65 s_{13} s_{23}+256 s_{23}^2\right)+131 s_{12} 
s_{13} s_{23} (s_{13}+s_{23})+140 s_{13}^2 s_{23}^2}{756 s_{12}^3 (3 
\epsilon -1)}
\nonumber\\ &
+\frac{-9 s_{12}^4+6 s_{12}^3 (s_{13}+s_{23})+3 s_{12}^2 
\left(7 s_{13}^2+s_{13} s_{23}+7 s_{23}^2\right)-s_{12} s_{13} s_{23} 
(s_{13}+s_{23})+2 s_{13}^2 s_{23}^2}{54 s_{12}^3} 
\nonumber\\ &
+\frac{\epsilon  
\left(s_{12}^3 (s_{13}+s_{23})+s_{12}^2 (s_{13}+s_{23})^2+s_{12} 
s_{13} s_{23} (s_{13}+s_{23})+s_{13}^2 s_{23}^2\right)}{9 
s_{12}^3}
\,.
\end{align}

For the corresponding form factor in ${\cal N}=4$ SYM, the coefficients can be given as:
\begin{align}
c_1^{{\cal N}=4} = & {1\over \epsilon^2} \,, \\
c_2^{{\cal N}=4} = & \frac{9}{s_{12} \epsilon ^3} -\frac{4 s_{12} s_{13}+4 s_{12} 
s_{23}+61 s_{13} s_{23}}{2 s_{12} s_{13} s_{23} \epsilon ^2}+\frac{6 
s_{12} s_{13}+6 s_{12} s_{23}+47 s_{13} s_{23}}{2 s_{12} s_{13} 
s_{23} \epsilon } 
\nonumber\\ &
+\frac{2}{s_{12} (\epsilon -1)}
-\frac{s_{12}^2+s_{13}^2+s_{23}^2-6 s_{13} s_{23}}{2 s_{12} s_{13} 
s_{23} (2 \epsilon -1)} +\frac{3 \left(s_{12}^2+s_{13}^2+s_{23}^2\right)}{2 s_{12} s_{13} s_{23}} \,, \\
c_3^{{\cal N}=4} = & \frac{s_{12}^2+s_{13}^2+s_{23}^2+2 s_{123}^2}{2 
s_{12} s_{13} s_{23} (2 \epsilon -1)}-\frac{3 \left(s_{12}^2+s_{13}^2+s_{23}^2+2 
s_{123}^2\right)}{2 s_{12} s_{13} 
s_{23}} \,,\\
c_4^{{\cal N}=4} = & \frac{3}{2 \epsilon ^2}-\frac{7}{2 \epsilon }-3 \,, \\
c_5^{{\cal N}=4} = & \frac{(s_{13}+s_{23})^2}{s_{13} s_{23} \epsilon } \,, \\
c_6^{{\cal N}=4} = & \frac{s_{23}}{\epsilon }-\frac{s_{23} (s_{12}-s_{13})}{s_{12} (3 \epsilon -1)} \,, \\
c_7^{{\cal N}=4} = & \frac{s_{123}^2-2 s_{13} s_{23}}{2 
s_{12} (2 \epsilon -1)}+\frac{s_{12}^2-s_{13}^2-s_{23}^2-3 s_{12} s_{13}-3 s_{12} 
s_{23}}{3 s_{12} (3 \epsilon 
-1)}-\frac{s_{12}^2-s_{13}^2-s_{23}^2}{6 s_{12}} \,.
\end{align}

We can see that the QCD coefficients are much more complicated comparing the ${\cal N}=4$ coefficients. However, interestingly, after the series expansion around $\epsilon = 0$, we find they are identical to $c_i$ up the  ${\cal O}(\epsilon^m)$ order indicated below:
\begin{align}
c_1 & =  {1\over \epsilon^2} + {\cal O}(\epsilon^1) \,, \\
c_2 & = \frac{9}{s_{12} \epsilon ^3}  + {\cal O}(\epsilon^{-2})  \,, \\
c_3 & = -\frac{2 \left(s_{12}^2+s_{13}^2+s_{23}^2+2 s_{123}^2\right)}{s_{12} \
s_{13} s_{23}} + {\cal O}(\epsilon^1) \,,\\
c_4 & = \frac{3}{2 \epsilon ^2} + {\cal O}(\epsilon^{-1}) \,, \\
c_5 & = \frac{(s_{13}+s_{23})^2}{s_{13} s_{23} \epsilon }   + {\cal O}(\epsilon^1)\,, \\
c_6 & = \frac{s_{23}}{\epsilon }+\frac{(s_{12}-s_{13}) 
s_{23}}{s_{12}}+\frac{3 (s_{12}-s_{13}) s_{23} \epsilon }{s_{12}}  + {\cal O}(\epsilon^2) \,, \\
c_7 & = -s_{12}+(-2 s_{12}+s_{13}+s_{23}) \epsilon +\frac{\left(-5 
s_{12}^2+s_{13}^2+s_{23}^2+5 s_{12} s_{13}+5 s_{12} s_{23}\right) 
\epsilon ^2}{s_{12}} + {\cal O}(\epsilon^3) \,.
\end{align}

\end{document}